\documentclass[sigconf]{acmart}

\usepackage[table]{xcolor}
\usepackage{enumitem}
\usepackage{algorithmic}
\usepackage[ruled, linesnumbered]{algorithm2e}
\usepackage{booktabs} % 提供 toprule, midrule
\usepackage{multirow} % 提供 multirow
\usepackage{graphicx} % 提供 resizebox
\usepackage{makecell} % 【关键】提供表头换行
\AtBeginDocument{%
  }

\author{Miaomiao Cai}
\affiliation{%
  \institution{National University of Singapore}
  \city{Singapore}
  \country{Singapore}
  }
\email{cmm.hfut@gmail.com}

\author{Yunshan Ma}
\affiliation{%
  \institution{Singapore Management University}
  \city{Singapore}
  \country{Singapore}
  }
\email{ysma@smu.edu.sg}

\author{Fangqi Zhu}
\affiliation{%
  \institution{Hefei University of Technology}
  \city{Hefei}
  \country{China}}
\email{2023212361@mail.hfut.edu.cn}

\author{Junfeng Fang}
\authornote{Corresponding Author}
\affiliation{%
  \institution{National University of Singapore}
  \city{Singapore}
  \country{Singapore}}
\email{fangjf1997@gmail.com}

\author{Zhijie Zhang}
\affiliation{%
  \institution{Hefei University of Technology}
  \city{Hefei}
  \country{China}}
\email{zhijiezhang021@gmail.com}

\author{Zhiyong Cheng}
\affiliation{%
  \institution{Hefei University of Technology}
  \city{Hefei}
  \country{China}}
\email{jason.zy.cheng@gmail.com}

\author{Xiang Wang}
\affiliation{%
  \institution{University of Science and Technology of China	}
  \city{Hefei}
  \country{China}}
\email{xiangwang1223@gmail.com}

\author{See-Kiong Ng}
\affiliation{%
  \institution{National University of Singapore}
  \city{Singapore}
  \country{Singapore}
  }
\email{seekiong@nus.edu.sg}

\copyrightyear{2026}
\acmYear{2026}
\setcopyright{cc}
\setcctype{by}
\acmConference[KDD '26]{Proceedings of the 32nd ACM SIGKDD Conference on Knowledge Discovery and Data Mining V.2}{August 09--13, 2026}{Jeju Island, Republic of Korea}
\acmBooktitle{Proceedings of the 32nd ACM SIGKDD Conference on Knowledge Discovery and Data Mining V.2 (KDD '26), August 09--13, 2026, Jeju Island, Republic of Korea}
\acmDOI{10.1145/3770855.3818191}
\acmISBN{979-8-4007-2259-2/2026/08}

\renewcommand{\shortauthors}{Miaomiao Cai et al.}

\begin{document}



\settopmatter{authorsperrow=4}

\title{Dynamic Spectral Denoising with Global-Context Attention for Multi-Behavior Recommendation}

\newcommand{\shortname}{\emph{\textbf{SpectraMB}}}

\newcommand{\fullname}{\textit{Dynamic \underline{\textbf{Spectra}}l De\-nois\-ing with Glob\-al-Con\-text At\-ten\-tion for \underline{\textbf{M}}ul\-ti-\underline{\textbf{B}}e\-hav\-ior Rec\-om\-men\-da\-tion (\textbf{SpectraMB})}}

\begin{abstract}

Multi-behavior recommendation improves target-behavior prediction by exploiting heterogeneous auxiliary feedback (e.g., \textit{view}, \textit{collect}, and \textit{cart}), yet its robustness is often undermined by behavior-dependent noise and inconsistency.
We argue that the key bottleneck is not merely noisy behaviors, but a \textit{representation-level failure} caused by two coupled heterogeneities.
First, \textbf{\textit{intra-behavior representation entanglement}} arises when multi-hop propagation blends incidental signals with true preferences in the embedding space. This entanglement renders coarse spatial denoising ineffective, since it cannot suppress noise without sacrificing weak-but-informative niche signals.
Second, \textbf{\textit{inter-behavior reliability heterogeneity}} complicates cross-behavior fusion, as the predictive value of auxiliary behaviors varies substantially across users and contexts. Without reliability calibration, aggregation can be dominated by frequent yet untrustworthy signals, leading to target-intent drift.
Existing methods typically address these issues in isolation and often fail when entanglement makes reliability estimation itself unstable.

To resolve this robustness bottleneck, we propose~\fullname, a target-oriented model that performs \emph{representation purification before reliability-aware fusion}.
To mitigate intra-behavior entanglement, ~\shortname~ introduces \textbf{Dynamic Feature-Level Spectral Filtering}, which re-parameterizes embeddings along the feature dimension into a feature-frequency space and learns view-adaptive spectral modulation end-to-end under target supervision, enabling component-wise purification without hand-crafted frequency assumptions.
Built on purified representations, ~\shortname~ further proposes \textbf{Global-Context Attention Fusion}, which uses the purified global representation as a stable context anchor to assess view compatibility and perform reliability-aware aggregation, while a residual global backbone preserves stable collaborative structure.
Extensive experiments on three real-world datasets show that ~\shortname~ achieves the best results in most evaluation settings and exhibits improved robustness under noisy interactions. 
Our implementation is available at \url{https://github.com/miaomiao-cai2/SpectraMB-KDD2026}.

\end{abstract}


\maketitle

\section{Introduction}

Recommender systems (RS) have become essential components of modern information services, helping users discover preferred items from massive candidate pools~\cite{chen2019efficient, Chen2020RevisitingGB,ma2024multimodal,yu2025fashiondpo,yu2024smart}.
Most practical recommenders optimize a \emph{target} behavior such as purchase, yet target feedback is often sparse, which limits the coverage and granularity of learned preferences~\cite{yang2023generative, Wu2020SelfsupervisedGL}.
To alleviate sparsity, Multi-Behavior Recommendation (MBR) leverages heterogeneous interaction signals (e.g., \textit{view}, \textit{collect}, and \textit{cart}) to enrich supervision and improve target-behavior prediction (e.g., \textit{purchase})~\cite{jin2020multi,cheng2023multi,meng2023parallel,yan2024behavior}.
However, user behaviors in real-world logs are inherently non-uniform and context-dependent.
They differ in intent strength, stability, and trustworthiness, and thus auxiliary behaviors can be noisy, ambiguous, or driven by temporary contexts rather than reflecting true target intent~\cite{han2024efficient,gao2022self,chuang2022robust,wang2025unleashing，wang2026trustworthy}.
As a result, naively exploiting multi-behavior signals may introduce noise and bias that distort preference modeling and degrade target-behavior recommendation performance~\cite{zhang2023denoising,han2024efficient,cai2025neighborhood,he2024denoising}.

Despite the effectiveness of multi-behavior learning, the major challenge remains \textit{how to exploit heterogeneous auxiliary feedback without being misled by its noise and inconsistency}.
To improve robustness, existing studies have developed a series of representation learning and aggregation strategies, many of which are built upon spatial message passing on behavior graphs and embedding-space fusion~\cite{he2024denoising,zhang2023denoising,han2024efficient,cai2025neighborhood,cai2025graph,DBLP:journals/tois/ZhuoQHDLW24}.
A representative line designs behavior-aware propagation and weighting mechanisms to reduce the impact of unreliable interactions during message passing~\cite{jin2020multi,yan2023cascading,cheng2023multi}.
Another line introduces self-supervised or contrastive objectives to enhance cross-behavior consistency, encouraging representations learned from different behaviors to share target-relevant semantics~\cite{gu2022self,xu2023multi,ma2025graph, wu2024multi}.
There are also disentanglement or projection-based designs that separate shared and behavior-specific factors to alleviate negative transfer across behaviors~\cite{meng2023parallel,yan2024behavior}.
More recently, invariant-learning inspired formulations treat different behaviors as heterogeneous environments and aim to distill behavior-stable preference factors for better generalization~\cite{yan2025user,wang2022causal, wang2026trustworthy}.

\begin{figure}

    \centering
    \includegraphics[width =\linewidth]{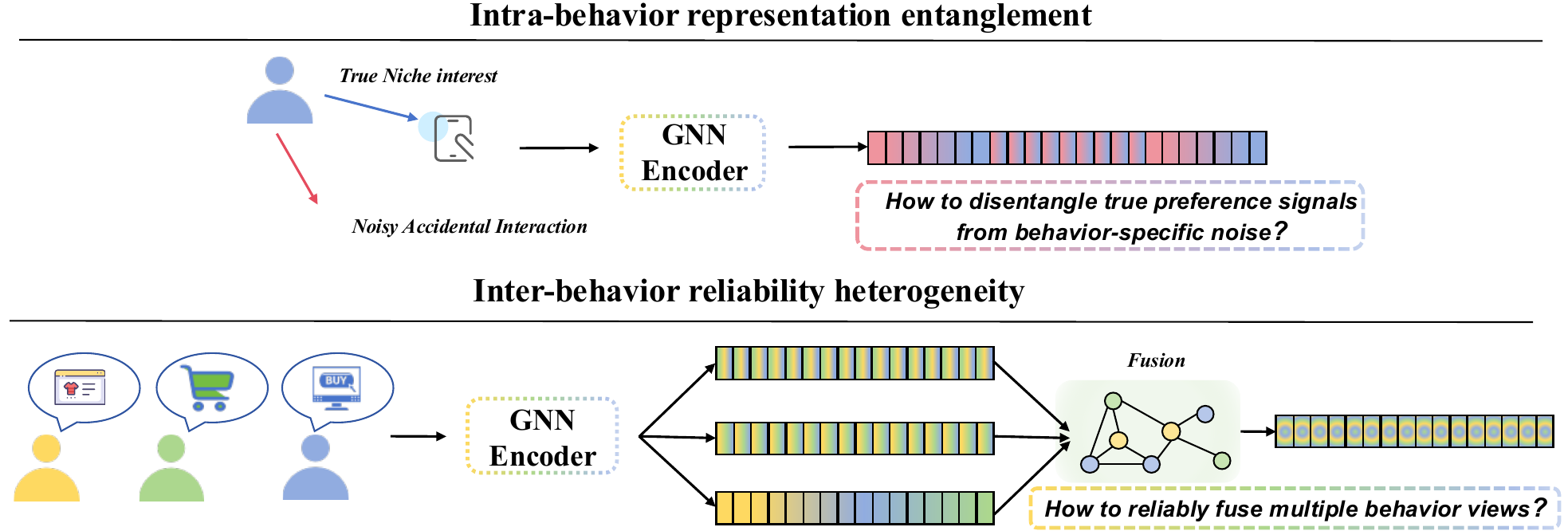}
    \caption{Illustration of representation entanglement and reliability heterogeneity.}
    \label{fig:intro}
    \vspace{-0.6cm}
\end{figure}

Despite these efforts, robust target-behavior learning in MBR remains a major challenge.
We argue that the main difficulty is not merely noisy auxiliary behaviors, but the representation entanglement induced by multi-hop propagation, through which behavior-dependent perturbations become inseparably coupled with true preference factors in the embedding space~\cite{He2020LightGCNSA,Yu2021AreGA}.
Once entanglement occurs, it becomes fundamentally difficult to distinguish noise from informative signals, which in turn complicates both denoising and multi-behavior fusion.
Two coupled heterogeneities lie at the core of this robustness bottleneck.

\begin{itemize}[leftmargin=*]

    \item \textbf{Intra-behavior representation entanglement.} Even within the same behavior (e.g., view), observed interactions may mix true preference with incidental actions~\cite{zhao2025symmetric}. 
    After multi-hop propagation on the behavior graph, incidental signals propagate together with preference evidence and become blended into the same latent dimensions of the learned embedding. 
    When such entanglement emerges in the embedding space, coarse spatial-domain treatments~\cite{he2024denoising,zhang2025dembr,han2024efficient,cai2025neighborhood}, including smoothing, pruning, or gating, often lack the granularity required to selectively suppress perturbations without weakening weak-but-informative niche signals. This limitation is further aggravated in the spatial domain, where a true but sparse niche interest and an accidental noisy interaction may exhibit highly similar topological patterns~\cite{oono2019graph,li2018deeper,wu2019simplifying}, since both can appear as weakly supported, isolated edges far from the user's major interaction cluster and are therefore hard to distinguish based solely on topology.

    \item \textbf{Inter-behavior reliability heterogeneity.} Auxiliary behaviors do not equally contribute to the target behavior, and their predictive reliability varies across users, contexts, and even item categories~\cite{wang2024distributionally,cai2024popularity}. 
    For some users, views may closely correlate with purchases, while for others they primarily reflect exploration. 
    When reliability is not assessed under the current context, aggregation can be dominated by frequent yet unreliable behaviors, which causes the fused preference to drift away from true target intent. 
    This issue is further coupled with intra-behavior entanglement, because once view representations are contaminated, reliability estimation becomes unstable~\cite{meng2023parallel,jin2020multi}, and fusion can still be misled even by sophisticated weighting mechanisms.

\end{itemize}

To address this robustness bottleneck caused by representation entanglement and reliability heterogeneity, we propose~\fullname, a target-oriented framework that performs \emph{representation purification before reliability-aware fusion}.
Specifically, ~\shortname~ introduces dynamic spectral filtering to suppress entangled perturbations within each behavior view, and utilizes a global-context~\cite{ding2021leveraging} attention mechanism to select trustworthy auxiliary signals based on their compatibility with a purified global preference structure.
First, to address intra-behavior representation entanglement, we introduce \emph{Dynamic Feature-Level Spectral Filtering}, which employs the Fast Fourier Transform (FFT) to efficiently capture global feature dependencies and map embeddings into a frequency basis where stochastic noise is physically separable from consistent preference trends.
Rather than further altering spatial message passing, we shift purification from the spatial domain to a \emph{feature-frequency} view.
We apply the Fast Fourier Transform (FFT)~\cite{cooley1965algorithm,lee2022fnet,xu2019frequency} along the embedding dimension to re-parameterize each behavior representation into a fixed basis, enabling different components to be modulated independently under target supervision.
This design enables selective suppression of incidental perturbations while preserving weak-but-informative preference signals.
Second, built upon purified representations, we introduce \emph{Global-Context Attention Fusion} to address inter-behavior reliability heterogeneity.
It uses the purified global representation as a stable context anchor to compute compatibility-based attention over behavior views, and combines the fused semantics with a residual global backbone to retain robust collaborative structure.
In this way, ~\shortname~ improves robustness within each behavior view and calibrates cross-behavior reliability under the target objective, yielding more reliable target preference modeling.
Extensive experiments on three real-world datasets demonstrate that ~\shortname~ achieves superior target-behavior ranking performance in most evaluation settings, with consistent improvements in NDCG across all datasets. 
Ablation and in-depth studies further confirm the roles of feature-frequency purification for mitigating intra-behavior entanglement and global-context fusion for handling inter-behavior reliability heterogeneity, including robustness evaluation under noisy settings.
The main contributions of this work are summarized as follows:

\begin{itemize}[leftmargin=*]

    \item We propose \textbf{Dynamic Feature-Level Spectral Filtering} to mitigate intra-behavior representation entanglement. By re-parame-terizing embeddings into a \emph{feature-frequency} space and learning view-adaptive spectral modulation under target supervision, it enables component-wise purification without hand-crafted frequency assumptions \textbf{while preserving weak signals}.

    \item We propose \textbf{Global-Context Attention Fusion} to address inter-behavior reliability heterogeneity.
    It uses the purified global representation as a context anchor to calibrate the reliability of each behavior view and aggregates them with reliability-aware attention, while a residual global backbone is retained to stabilize collaborative structure for target prediction.

    \item Extensive experiments on three datasets demonstrate that ~\shortname~ achieves the best results in most evaluation settings and consistently improves ranking quality measured by NDCG. Ablation and robustness analyses further verify the effectiveness of both modules, especially under noisy interactions. 

\end{itemize}

\section{Related Work}

\subsection{Multi-Behavior Recommendation}

Multi-behavior recommendation (MBR) alleviates target-feedback sparsity by leveraging heterogeneous auxiliary interactions to enhance target behavior prediction~\cite{gao2019learning,jin2020multi,cheng2023multi,yan2024behavior}.
Early studies extend collaborative filtering to jointly model multiple interaction types and transfer supervision across behaviors~\cite{gao2019learning,Rendle2009BPRBP}.
With the rise of graph representation learning, graph-based MBR has become the dominant paradigm, where behavior-specific user--item graphs are encoded via message passing and then fused for target recommendation~\cite{jin2020multi,cheng2023multi,yan2023cascading}.
Representative methods perform within-behavior propagation followed by behavior-aware aggregation~\cite{jin2020multi}, or model inter-behavior dependency through cascading architectures to progressively enhance target modeling with auxiliary signals~\cite{yan2023cascading,cheng2023multi}.

Despite their effectiveness, most MBR approaches operate in the spatial domain with view-level fusion.
When auxiliary signals are repeatedly propagated through multi-hop neighborhoods, incidental interactions may be reinforced and become entangled with genuine preference evidence in the learned embeddings~\cite{li2018deeper,wu2019simplifying,oono2019graph}, which undermines robust target-oriented modeling.

\subsection{Denoising and Robust Learning for MBR}

To mitigate noisy and heterogeneous auxiliary feedback, recent studies introduce denoising and robust learning strategies for MBR~\cite{zhang2023denoising,han2024efficient,he2024denoising,cai2025neighborhood,wang2025unleashing}.
One direction employs self-supervised or contrastive objectives to enhance cross-behavior consistency and improve robustness under sparse supervision, such as SGL-style or graph contrastive extensions for MBR~\cite{gu2022self,xu2023multi,ma2025graph,wu2024multi,zhao2025symmetric}.
Another line explores structured modeling such as disentanglement or projection-based designs to separate shared and behavior-specific factors, as exemplified by PKEF and behavior-contextualized modeling frameworks~\cite{meng2023parallel,yan2024behavior}.
Several approaches also perform explicit filtering or pruning to suppress noisy interactions, including prompt-/filter-based denoising and memory pruning mechanisms adopted in  DeMBR~\cite{zhang2023denoising,han2024efficient,zhang2025dembr}.
More recently, invariant-learning perspectives such as UIPL have been introduced to extract behavior-stable preference patterns under heterogeneity and distribution shifts~\cite{yan2025user,arjovsky2019invariant,ahuja2020invariant,krueger2021out} \textbf{for better generalization}.

However, these methods primarily operate on spatial representations or behavior-level aggregation after multi-hop propagation~\cite{cai2026rmbrec}.
Once noise and preference cues become entangled in the embedding space, spatial aggregation, view-level regularization, or local filtering may still struggle to selectively suppress perturbations without sacrificing weak-but-informative niche signals.
Motivated by recent advances that leverage frequency-domain perspectives for representation learning and filtering~\cite{cooley1965algorithm,xu2019frequency,lee2022fnet,peng2022less,zhang2023contrastive,fan2024hierarchical,ong2025spectrum}, we revisit MBR from a feature-frequency view and introduce an adaptive purification mechanism tailored to multi-behavior interactions \textbf{for robust target prediction}.

\section{Model}

\begin{figure*}[t]
    \centering
    \includegraphics[width=0.92\linewidth]{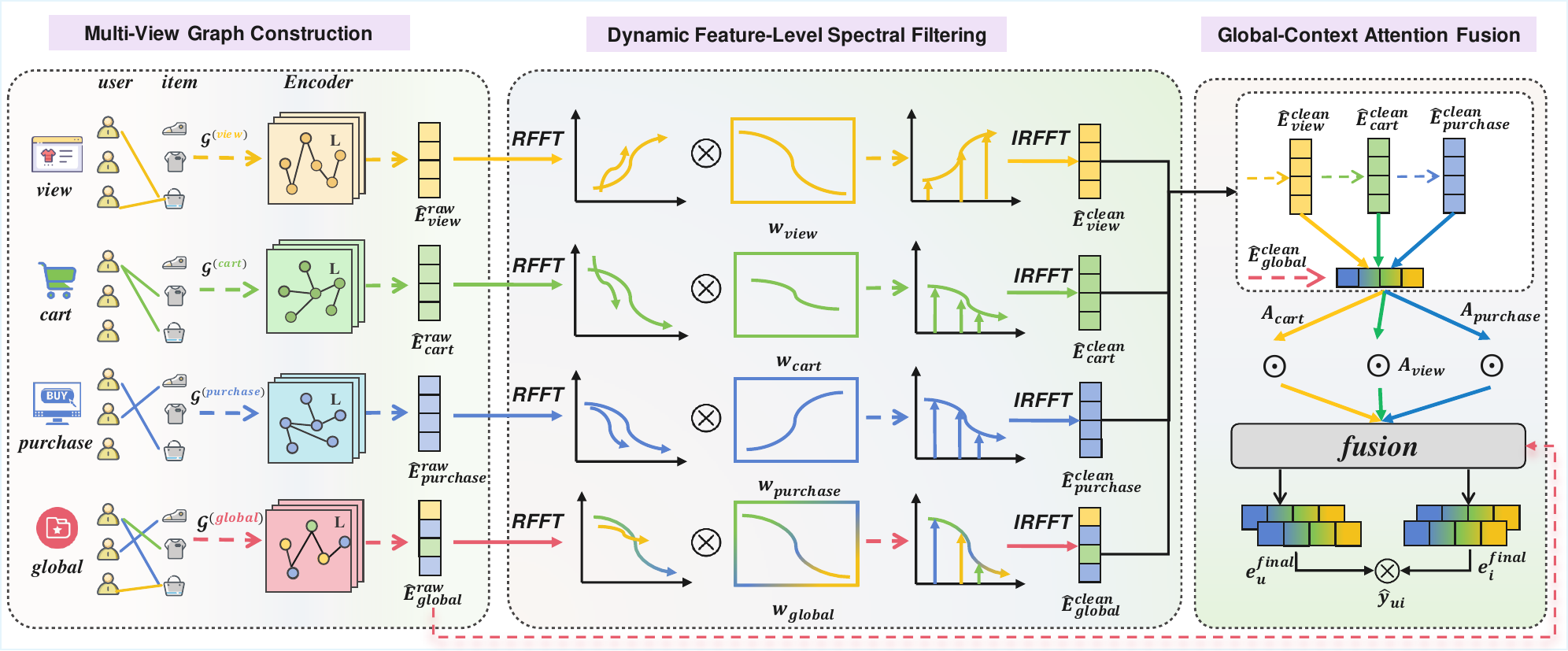}
    \caption{Overview of ~\shortname. Spectral purification and global-context attention for robust multi-behavior recommendation.}
    \label{fig:framework}
    \vspace{-0.5cm}
\end{figure*}

As illustrated in Figure~\ref{fig:framework}, we propose ~\shortname~ to address two coupled heterogeneities in multi-behavior recommendation: \textit{intra-behavior representation entanglement} and \textit{inter-behavior reliability heterogeneity}.
Our core principle is \emph{representation purification before reliability-aware fusion}.
Specifically, multi-hop propagation on each behavior view can couple behavior-dependent perturbations with true preference factors in the embedding space, which makes subsequent denoising and fusion fundamentally harder.
Accordingly, ~\shortname~ first introduces \textit{Dynamic Feature-Level Spectral Filtering}, which shifts purification from the spatial domain to a \textit{feature-frequency perspective} by re-parameterizing each view embedding and learning view-adaptive spectral modulation under target supervision.
Built on purified representations, ~\shortname~ further introduces \textit{Global-Context Attention Fusion}, which uses the purified global representation as a context anchor to calibrate cross-behavior reliability and prevents frequent but unreliable auxiliary behaviors from dominating the fused preference representation.

\subsection{Multi-View Graph Construction}

Let $\mathcal{U}=\{u_1,\dots,u_N\}$ and $\mathcal{I}=\{i_1,\dots,i_M\}$ denote the user and item sets, respectively.
In a multi-behavior setting, user--item interactions are associated with a behavior set $\mathcal{B}=\mathcal{B}_{aux}\cup\{target\}$, where $\mathcal{B}_{aux}$ contains auxiliary behaviors (e.g., \textit{view}, \textit{cart}) and $target$ denotes the target behavior (e.g., \textit{purchase}).
For each $b\in\mathcal{B}$, we encode implicit feedback as a binary matrix $\mathbf{Y}_b\in\{0,1\}^{N\times M}$, where $y_{ui,b}=1$ indicates that user $u$ performs behavior $b$ on item $i$.

To retain behavior-specific interaction semantics~\cite{jin2020multi,cheng2023multi,yan2023cascading,yan2024behavior}, we construct a behavior-specific user--item bipartite graph $\mathcal{G}_b=(\mathcal{V},\mathcal{E}_b)$ for each $b\in\mathcal{B}$, where $\mathcal{V}=\mathcal{U}\cup\mathcal{I}$ and $\mathcal{E}_b=\{(u,i)\mid y_{ui,b}=1\}$.
Meanwhile, since each individual behavior can be sparse and noisy in realistic logs~\cite{zhang2023denoising,han2024efficient,cai2025neighborhood,wang2025unleashing}, we additionally build a unified global interaction graph $\mathcal{G}_{global}=(\mathcal{V},\mathcal{E}_{global})$ by aggregating all observed interactions:
\begin{equation}
    \mathcal{E}_{global}=\{(u,i)\mid \exists b\in\mathcal{B},~ y_{ui,b}=1\}.
\end{equation}
In other words, an edge $(u,i)$ is included in the global view whenever \emph{any} behavior between $u$ and $i$ is observed.
This global view provides a holistic interaction context that is typically more stable than any single behavior view, and it will later serve as a context anchor for reliability-aware fusion.
Unless otherwise stated, all graphs are constructed using training interactions only to avoid using future information.

Building on these multi-view graphs, we adopt LightGCN~\cite{He2020LightGCNSA} as the backbone encoder due to its effectiveness and efficiency in capturing high-order collaborative signals with simplified neighborhood aggregation~\cite{He2020LightGCNSA,Wang2019NeuralGC}.
We initialize trainable embeddings for all nodes as $\mathbf{E}^{(0)}\in\mathbb{R}^{(N+M)\times d}$, where $d$ is the embedding dimension.
For a given graph view $\mathcal{G}$ (either $\mathcal{G}_b$ or $\mathcal{G}_{global}$), LightGCN propagates embeddings via normalized neighbor aggregation~\cite{He2020LightGCNSA}.
At layer $l$, the updates are:
\begin{equation}
    \mathbf{e}_u^{(l+1)}=\sum_{i\in\mathcal{N}_u}\frac{1}{\sqrt{|\mathcal{N}_u||\mathcal{N}_i|}}\mathbf{e}_i^{(l)}, \quad
    \mathbf{e}_i^{(l+1)}=\sum_{u\in\mathcal{N}_i}\frac{1}{\sqrt{|\mathcal{N}_i||\mathcal{N}_u|}}\mathbf{e}_u^{(l)},
    \label{eq:gcn_prop}
\end{equation}
where $\mathcal{N}_u$ and $\mathcal{N}_i$ denote the neighbor sets of $u$ and $i$ in $\mathcal{G}$.
After stacking $L$ layers, we average embeddings across layers to obtain the spatial representation:
\begin{equation}
    \mathbf{E}^{raw}=\frac{1}{L+1}\sum_{l=0}^{L}\mathbf{E}^{(l)}.
    \label{eq:gcn_readout}
\end{equation}

Applying the same LightGCN encoder~\cite{He2020LightGCNSA} (\emph{with shared parameters} across views) to the global view and each behavior view yields the raw global representation $\mathbf{E}^{raw}_{global}$ and behavior-specific representations $\{\mathbf{E}^{raw}_b\}_{b\in\mathcal{B}}$.
These spatial representations capture high-order collaborative dependencies within each view~\cite{He2020LightGCNSA}, but multi-layer propagation can also couple behavior-dependent perturbations with preference factors in the latent space~\cite{li2018deeper,wu2019simplifying,oono2019graph}.
This motivates our subsequent Dynamic Feature-Level Spectral Filtering module, which purifies entangled components in a feature-frequency domain before reliability-aware multi-view fusion.

\subsection{Dynamic Feature-Level Spectral Filtering}

In multi-behavior recommendation, \textit{intra-behavior representation entanglement} arises from behavior-dependent noise, which varies across interaction types~\cite{zhang2023denoising,han2024efficient}. Auxiliary behaviors like \textit{view} and \textit{cart} often include incidental signals, while even the target behavior may have preference-irrelevant cases (e.g., gifting). These noise components propagate and mix with genuine signals in multi-hop aggregation, making noise inseparable from preferences~\cite{li2018deeper,wu2019simplifying,oono2019graph}.
Spatial GNN encoders~\cite{He2020LightGCNSA, Wang2019NeuralGC} capture high-order dependencies, but their uniform aggregation of embeddings fails to distinguish noise from preference signals~\cite{li2018deeper, wu2019simplifying}. This results in over-smoothing or under-smoothing, which weakens weak but informative signals—critical in multi-behavior recommendation, especially under target sparsity.~\cite{zhang2023denoising, han2024efficient, zhang2025dembr, cai2025neighborhood}.

To address this, we shift to a \textit{feature-frequency} perspective, applying the Fast Fourier Transform (FFT) along the embedding dimension~\cite{cooley1965algorithm,lee2022fnet,xu2019frequency,ong2025spectrum,fan2024hierarchical,zhang2023contrastive,peng2022less}. Unlike graph-based spectral filtering, which is costly and graph-dependent~\cite{peng2022less,lu2024addressing}, our method uses FFT’s reversible transformation to expose and modulate spectral components without assuming specific frequencies are noisy~\cite{lee2022fnet, xu2019frequency}. This graph-agnostic approach enables adaptive, target-supervised modulation.
We treat each node's embedding $\mathbf{e} \in \mathbb{R}^d$ as a real-valued signal along the feature dimension. Using the \textbf{Real-valued Fast Fourier Transform (RFFT)}, we retain non-redundant frequency components for efficient purification and modulation of relevant signals \textbf{for robust preference learning}.

Specifically, for each node, we treat its spatial embedding $\mathbf{e}\in\mathbb{R}^d$ as a discrete real-valued signal defined along the feature dimension.
Since the embeddings are real-valued, their Fourier spectrum exhibits \textit{conjugate symmetry}.
Leveraging this property, we apply the \textbf{Real-valued Fast Fourier Transform (RFFT)} to keep only the non-redundant frequency components:
\begin{equation}
    \label{eq:rfft}
    \hat{\mathbf{E}}^{raw}_{global} = \text{RFFT}(\mathbf{E}^{raw}_{global}),
    \quad
    \hat{\mathbf{E}}^{raw}_{b} = \text{RFFT}(\mathbf{E}^{raw}_{b}),
\end{equation}
where $\widehat{\mathbf{E}}_{\mathrm{global}}^{\mathrm{raw}}, \widehat{\mathbf{E}}_{b}^{\mathrm{raw}} \in \mathbb{C}^{(N+M) \times d_f}$ are the complex-valued feature-frequency representations transformed along the embedding dimension, and $d_f = \lfloor d/2 \rfloor + 1$ denotes the number of non-redundant frequency bins retained by RFFT for real-valued inputs. Note that $d_f$ is not a tunable hyperparameter; it is determined by the Hermitian symmetry of the Fourier spectrum of real-valued embeddings. Retaining fewer frequency bins independently of $d$ would introduce an additional truncation operation and would no longer permit exact reconstruction of the original raw representation through IRFFT.

Unlike explicit content features in text or images, ID-based embeddings in MBR do not assign predefined human-interpretable semantics to individual latent dimensions. Accordingly, applying RFFT along the embedding dimension serves as a reversible basis transformation before spectral modulation, rather than destroying explicit semantic priors. We therefore impose no fixed interpretation on specific frequency ranges; instead, the feature-frequency representation provides a component-wise parameterization through which the target objective learns which components should be suppressed or emphasized. 

Motivated by the heterogeneous perturbations across behavior views, we propose a \textit{\textbf{Behavior-Adaptive Spectral Filter}} that learns view-specific modulation strategies for different interaction types.
Specifically, we introduce independent learnable spectral modulators, $\mathbf{W}_{global} \in \mathbb{R}^{d_f \times d_f}$ for the global view, and separate matrices $\{\mathbf{W}_b\}_{b \in \mathcal{B}}$ for each behavior-specific view, where each $\mathbf{W}$ is constrained to be diagonal for efficient component-wise modulation.
Filtering is conducted via a linear transformation directly in the feature-frequency domain:
\begin{equation}
    \hat{\mathbf{E}}_{global}^{clean} = \hat{\mathbf{E}}_{global}^{raw} \mathbf{W}_{global}, \quad
    \hat{\mathbf{E}}_b^{clean} = \hat{\mathbf{E}}_b^{raw} \mathbf{W}_b.
    \label{eq:spec_mod}
\end{equation}
Here, $\mathbf{W}_{global}$ and $\{\mathbf{W}_b\}_{b\in\mathcal{B}}$ act as learnable \emph{spectral modulators} in the feature-frequency domain.
They provide fine-grained, view-adaptive control over embedding components by reweighting frequency bins of $\hat{\mathbf{E}}^{raw}_{global}$ and $\hat{\mathbf{E}}^{raw}_b$.
Crucially, the modulation is learned end-to-end under the \emph{target} ranking supervision, such that gradients from the target objective backpropagate to $\mathbf{W}_{global}$ and $\mathbf{W}_b$, enabling the model to suppress components dominated by stochastic perturbations while emphasizing those that are most predictive of the target behavior.
As a result, ~\shortname~ avoids heuristic cutoff rules and can preserve informative components that are beneficial for niche preferences.

Finally, we reconstruct purified embeddings back to the spatial feature space via the \emph{Inverse Real-valued Fast Fourier Transform (IRFFT)}:
\begin{equation}
    \label{eq:irfft}
    \mathbf{E}^{clean}_{global} = \text{IRFFT}(\hat{\mathbf{E}}_{global}^{clean}), \quad
    \mathbf{E}^{clean}_b = \text{IRFFT}(\hat{\mathbf{E}}_b^{clean}).
\end{equation}
By performing view-adaptive spectral modulation in the feature-frequency domain under target supervision, ~\shortname~ enables component-wise purification beyond spatial-only aggregation, mitigating intra-behavior entanglement while preserving weak-but-informative signals for robust target behavior prediction.

\subsection{Global-Context Attention Fusion}

After Dynamic Feature-Level Spectral Filtering, we obtain purified representations $\mathbf{E}^{clean}_{global}$ and $\{\mathbf{E}^{clean}_b\}_{b \in \mathcal{B}}$.
However, multi-behavior fusion remains challenging due to \textit{inter-behavior reliability heterogeneity}, i.e., the predictive value of auxiliary behaviors varies substantially across users and contexts~\cite{jin2020multi,yan2024behavior,meng2023parallel}.
For example, \textit{view} may indicate strong purchase intent for some users but remain largely exploratory for others, and the behavior--target relation can also differ across item categories~\cite{yan2024behavior,meng2023parallel}.
Consequently, static fusion strategies (e.g., fixed weights, simple summation, or concatenation) are often quantity-driven and may allow frequent but unreliable auxiliary behaviors to dominate the fused representation, leading to target-intent drift~\cite{jin2020multi,yan2023cascading,cheng2023multi}.
This issue is further coupled with intra-behavior entanglement, because once view representations are contaminated, reliability estimation becomes unstable~\cite{meng2023parallel,jin2020multi}, and fusion can still be misled even by sophisticated weighting mechanisms.

To address this issue, we propose \textbf{Global-Context Attention Fusion}, which performs reliability-aware aggregation guided by a global context anchor.
Our key idea is to use the purified global representation $\mathbf{E}^{clean}_{global}$ as a stable context summary of user interests, and to assess each behavior view by its \textit{compatibility} with this global context.
This compatibility serves as a proxy for the reliability of a behavior signal under the current context, allowing the model to down-weight behavior views that deviate from the overall preference profile while emphasizing those that are more trustworthy for target preference prediction.
We use $\mathbf{E}^{clean}_{global}$ for reliability estimation, while preserving $\mathbf{E}^{raw}_{global}$ as a structural residual to stabilize the collaborative geometry.

Specifically, for each behavior $b \in \mathcal{B}$, we compute attention weights by explicitly measuring the compatibility between the purified behavior-specific representation and the purified global context.
For each behavior $b$, we first compute a node-wise compatibility matrix:
\begin{equation}
    \label{eq:compat}
    S_{b} = \tanh\left( E_{global}^{clean} W_{g}^{(b)} + E_{b}^{clean} W_{b}^{(b)} + v^{(b)} \right),
\end{equation}
where $W_{g}^{(b)}, W_{b}^{(b)} \in \mathbb{R}^{d \times d}$ are learnable weight matrices, and $v^{(b)} \in \mathbb{R}^{d}$ is the behavior-specific bias vector. Note that the addition of $v^{(b)}$ is performed via broadcasting along the node dimension to ensure dimensional consistency.
As a result, $\mathbf{S}_b\in\mathbb{R}^{|\mathcal{V}|\times d}$ provides a $d$-dimensional compatibility representation for each node, indicating how the behavior-specific embedding aligns with the global context at the representation level.

We then normalize $\{\mathbf{S}_b\}_{b\in\mathcal{B}}$ across behaviors for each node and each embedding dimension to obtain the attention matrices:
\begin{equation}
\label{eq:attn}
(\mathbf{A}_b)_{v,k}
=
\frac{\exp\!\left((\mathbf{S}_b)_{v,k}\right)}
{\sum_{b' \in \mathcal{B}} \exp\!\left((\mathbf{S}_{b'})_{v,k}\right)},
\quad
\forall v \in \mathcal{V},\; k=1,\ldots,d,
\end{equation}
where $\mathbf{A}_b\in\mathbb{R}^{|\mathcal{V}|\times d}$ denotes the normalized attention matrix for behavior $b$.
This node-wise and dimension-wise normalization enforces cross-behavior competition for each semantic dimension, encouraging reliability-driven contributions rather than frequency-driven dominance.

Finally, we integrate the reliability-aware fusion with a stable structural backbone through a weighted residual design~\cite{Chen2020RevisitingGB}.
Specifically, given the normalized attention matrices $\{\mathbf{A}_b\}_{b\in\mathcal{B}}$ and the purified behavior-specific representations $\{\mathbf{E}^{clean}_b\}_{b\in\mathcal{B}}$, we obtain the final node representations as:
\begin{equation}
\label{eq:fusion}
\mathbf{E}^{final}
=
\sum_{b\in\mathcal{B}}
\mathbf{A}_b \odot \mathbf{E}^{clean}_b
\;+\;
\alpha\,\mathbf{E}^{raw}_{global},
\end{equation}
where $\alpha$ controls the contribution of the global spatial backbone.
The attention-based term aggregates behavior views according to their context compatibility, yielding a reliability-aware representation for target preference prediction.
Meanwhile, injecting $\alpha\,\mathbf{E}^{raw}_{global}$ preserves the collaborative topology distilled by spatial propagation, stabilizing the embedding space and preventing the fusion from overly distorting structural signals.
Overall, this fusion strategy enables ~\shortname~ to jointly leverage reliable multi-behavior evidence and robust global structural signals for effective target behavior prediction.

\subsection{Prediction and Optimization}

After the above stages, we obtain the final representation matrix $\mathbf{E}^{final}\in\mathbb{R}^{|\mathcal{V}|\times d}$ for all nodes, where $\mathcal{V}=\mathcal{U}\cup\mathcal{I}$.
For a user $u\in\mathcal{U}$ and an item $i\in\mathcal{I}$, we retrieve their embeddings $\mathbf{e}^{final}_{u}$ and $\mathbf{e}^{final}_{i}$ from $\mathbf{E}^{final}$.
These embeddings jointly encode high-order collaborative structure captured by global spatial propagation and reliability-aware multi-behavior semantics learned through feature-frequency purification and global-context fusion.

To quantify the preference of user $u$ over item $i$ under the target behavior, we adopt the standard inner-product predictor~\cite{He2020LightGCNSA}:
\begin{equation}
    \label{eq:pred}
    \hat{y}_{ui} = {\mathbf{e}_u^{final}}^\top \mathbf{e}_i^{final},
\end{equation}
where $\hat{y}_{ui}$ is the predicted ranking score indicating the propensity that user $u$ will perform the target action (e.g., purchase) on item $i$.

Since our goal is to improve ranking performance on the \emph{target} behavior, we optimize ~\shortname~ using \textbf{only target-behavior supervision}.
Specifically, we employ the \textbf{Bayesian Personalized Ranking (BPR) loss}~\cite{Rendle2009BPRBP} and construct the training set $\mathcal{O}$ using target interactions.
Each training instance is a triplet $(u,i,j)$ where user $u$ has an observed target interaction with item $i$ (i.e., $y_{ui,target}=1$), and $j$ is a negative item sampled with no observed target interaction (i.e., $y_{uj,target}=0$)~\cite{ma2026negative}.
The objective is:
\begin{equation}
    \label{eq:bpr}
    \mathcal{L} = \sum_{(u, i, j) \in \mathcal{O}} - \ln \sigma(\hat{y}_{ui} - \hat{y}_{uj}) + \lambda \|\Theta\|_2^2,
\end{equation}
where $\sigma(\cdot)$ is the sigmoid function, $\Theta$ denotes all trainable parameters, and $\lambda$ controls $L_2$ regularization.

We emphasize that optimizing solely on the target behavior is a fundamental design principle of ~\shortname, distinguishing it from traditional Multi-Task Learning (MTL) approaches. In MTL, auxiliary behaviors are treated as ground-truth labels to be fitted. However, in our hypothesis, auxiliary interactions (e.g., views) inherently contain substantial noise and false positives—users often view without genuine intent~\cite{zhang2023denoising,han2024efficient}.
Including an auxiliary loss would fundamentally contradict our denoising objective, as it would explicitly force the model to encode these stochastic noise patterns into the representation. Instead, ~\shortname~ treats auxiliary behaviors as \textit{contextual evidence} rather than \textit{optimization targets}. By backpropagating gradients \textbf{only} from the target behavior loss, the spectral filters and attention modules are guided to extract only the information that is \textit{predictive of the target}, while automatically discarding the noise that is irrelevant to the final conversion. This ensures that the auxiliary data serves to refine, rather than distract, the target preference learning. 

\subsection{Complexity Analysis}

We first analyze the asymptotic complexity of ~\shortname. Let $|\mathcal{E}_{\Sigma}| = |\mathcal{E}_{\mathrm{global}}| + \sum_{b \in \mathcal{B}} |\mathcal{E}_{b}|$ denote the total number of edges across the global and behavior-specific graph views, and let $K = |\mathcal{B}| + 1$ denote the number of views. With $L$ propagation layers and embedding dimension $d$, the multi-view LightGCN encoder requires $O(L \cdot |\mathcal{E}_{\Sigma}| \cdot d)$ operations. Dynamic Feature-Level Spectral Filtering applies RFFT and IRFFT to all view representations and performs diagonal spectral modulation, leading to an additional cost of $O(K \cdot |\mathcal{V}| \cdot d \log d)$. Global-Context Attention Fusion further introduces $O(|\mathcal{B}| \cdot |\mathcal{V}| \cdot d^{2})$ operations for behavior-specific compatibility estimation. Therefore, the overall complexity of ~\shortname~ is $O(L \cdot |\mathcal{E}_{\Sigma}| \cdot d + K \cdot |\mathcal{V}| \cdot d \log d + |\mathcal{B}| \cdot |\mathcal{V}| \cdot d^{2})$. Since both $d$ and the number of behavior views are typically small in practical recommender systems, the additional node-wise spectral filtering and attention operations scale linearly with the number of nodes. Moreover, unlike exact Laplacian-based graph spectral filtering, our feature-level RFFT operation does not require eigendecomposition over the large interaction graph.

To further evaluate practical efficiency, we report the runtime comparison on the TMall dataset in Table~\ref{tab:runtime}. Compared with LightGCN, ~\shortname~ introduces additional per-epoch computation due to multi-view spectral purification and global-context attention fusion. Nevertheless, it converges in only 27 epochs, fewer than all compared methods, and its per-epoch training time is substantially lower than those of SimGCL and BIPN. While ~\shortname~ incurs a higher per-epoch testing cost, these results indicate that its additional robustness-oriented components provide a reasonable efficiency--effectiveness trade-off in practice.

\begin{table}[t]
\centering
\caption{Runtime comparison on the TMall dataset.}
\label{tab:runtime}
\small
\setlength{\tabcolsep}{4pt}
\begin{tabular}{lccc}
\toprule
\textbf{Model} & \textbf{Train / Epoch (s)} & \textbf{Test / Epoch (s)} & \textbf{Conv. Epochs} \\
\midrule
LightGCN    & 5.38    & 45.54 & 66  \\
SimGCL      & 784.63  & 15.71 & 200 \\
BIPN        & 1041.28 & 9.51  & 39  \\
UIPL        & 50.29   & 19.52 & 106 \\
\textbf{\shortname} & 110.71  & 56.25 & \textbf{27} \\
\bottomrule
\end{tabular}
\vspace{-0.3cm}
\end{table}

\section{Experiments}

\subsection{Experiment Settings}

\subsubsection{Datasets.}

\begin{table}[t]
    % \vspace{-0.5cm}
    \caption{Statistics of the three multi-behavior datasets.}
    \label{tab:dataset}
    \centering
    \resizebox{\linewidth}{!}{
    \begin{tabular}{lcccccc}
    \toprule
    % --- 上半部分表头 ---
    \textbf{Dataset} & \textbf{\#Users} & \textbf{\#Items} & \textbf{\#View} & \textbf{\#Collect} & \textbf{\#Cart}  & \textbf{\#Purchase} \\
    \midrule
    Taobao & 48,749 & 39,493 & 1,548,162 & - & 193,747 & 259,771 \\
    TMall & 41,738 & 11,953 & 1,813,498 & 221,514 & 1,996 & 287,158 \\
    \midrule
    \midrule
    % --- 下半部分表头 (重新定义列名) ---
    \textbf{Dataset} & \textbf{\#Users} & \textbf{\#Items} &  \textbf{\#Dislike} & \textbf{\#Neutral} & \multicolumn{2}{c}{\textbf{\#Like}} \\
    \midrule
    MovieLens-10M & 67,788 & 8,704 & 1,370,897 & 3,580,155 & \multicolumn{2}{c}{4,970,984}\\
    \bottomrule
    \end{tabular}
    }
    \vspace{-0.5cm}
\end{table}

We conduct experiments on three widely used real-world multi-behavior datasets: \textbf{Taobao}, \textbf{TMall}, and \textbf{MovieLens-10M}.
Taobao and TMall are real-world e-commerce logs that contain multiple types of user-item interactions, where \textit{purchase} is treated as the \textbf{target behavior} and the remaining behaviors (e.g., \textit{view}, \textit{collect}, and \textit{add-to-cart}) are used as auxiliary signals.
For \textbf{MovieLens-10M}, the original feedback is explicit ratings from 1 to 5.
Following common practice in prior work~\cite{yan2024behavior}, we convert ratings into three implicit behaviors, i.e., \textit{dislike}, \textit{neutral}, and \textit{like}, and treat \textit{like} as the \textbf{target behavior}.
Specifically, we map low ratings to \textit{dislike}, mid ratings to \textit{neutral}, and high ratings to \textit{like}.
This conversion enables a consistent multi-behavior learning setup on rating-based data.
Following standard preprocessing protocols~\cite{jin2020multi,cheng2023multi}, we convert logs into binary implicit feedback and remove duplicates.
When multiple interactions of the same behavior occur between a user and an item, we keep only the earliest occurrence to construct the adjacency matrices.
The dataset statistics are summarized in Table~\ref{tab:dataset}.

\subsubsection{Baselines.}

To demonstrate the effectiveness and robustness of ~\shortname, we compare it with a broad set of state-of-the-art baselines, including both single-behavior collaborative filtering methods and multi-behavior recommendation models.
(1) \textbf{General Collaborative Filtering.} We include representative single-behavior baselines that model only the target behavior (purchase), including \textbf{LightGCN}~\cite{He2020LightGCNSA} and the robust contrastive baseline \textbf{SimGCL}\\ ~\cite{Yu2021AreGA}.
(2) \textbf{Multi-Behavior Recommendation.} We further compare against representative MBR methods that leverage auxiliary behaviors for alleviating target sparsity, covering different modeling paradigms for behavior fusion and heterogeneity handling. Specifically, we include graph-based fusion models (\textbf{MBGCN}~\cite{jin2020multi}, \textbf{MB-CGCN}~\cite{yan2023cascading}), representation disentanglement and projection methods (\textbf{PKEF}~\cite{meng2023parallel}, \textbf{MISSL}~\cite{xu2023multi}), contrastive alignment approaches (\textbf{S-MBRec}~\cite{gu2022self}), and robust learning or denoising-oriented methods (\textbf{DeMBR}~\cite{zhang2025dembr}, \textbf{UIPL}~\cite{yan2025user}, \textbf{BCIPM}~\cite{zhang2023denoising}, \textbf{MBID}~\cite{xu2025multi}). 

% Detailed descriptions are provided in Appendix~\ref{app:baselines}.

\subsubsection{Evaluation and Implementation Details}

We use two widely adopted ranking metrics, \textbf{HR@$K$} and \textbf{NDCG@$K$}~\cite{herlocker2004evaluating}, with $K \in \{10, 20\}$.
We follow the \textbf{all-ranking} evaluation protocol, ranking the ground-truth item against all non-interacted items in the candidate set to avoid sampling bias.

We implement ~\shortname~ and all baselines using the PyTorch framework~\cite{paszke2019pytorch} and optimize the parameters using the Adam optimizer~\cite{kingma2014adam}.
For fair comparison, we set the embedding dimension to $d=64$ and use $L=2$ propagation layers for all GCN-based models.
We tune hyperparameters for baselines using their official implementations. For ~\shortname, we tune the residual coefficient $\alpha$ via grid search over $[0.1, 2.0]$ with a step size of $0.1$. All runtime results are measured under the same hardware environment
and stopping criterion.

\subsection{Overall Performance}

\begin{table*}[t]
\caption{Overall performance comparison with state-of-the-art methods. The best and second-best results are highlighted in \textbf{bold} and \underline{underlined}, respectively. ``Imp.'' denotes the relative change of ~\shortname~ over the strongest baseline. The asterisk ($*$) indicates that the improvement is statistically significant with $p < 0.05$.}
\label{tab:main_result}
\centering
\small
\resizebox{\linewidth}{!}{
\begin{tabular}{l|l|cc|ccccccccc|cc}
\toprule
\multirow{2.5}{*}{\textbf{Dataset}} & \multirow{2.5}{*}{\textbf{Metric}} 
& \multicolumn{2}{c|}{\textbf{Single-Behavior}} 
& \multicolumn{9}{c|}{\textbf{Multi-Behavior}} 
& \multicolumn{2}{c}{\textbf{Ours}} \\
\cmidrule(lr){3-4} \cmidrule(lr){5-13} \cmidrule(l){14-15}

& 
& \makecell[c]{\textbf{LightGCN} \\ \footnotesize (SIGIR'20)} 
& \makecell[c]{\textbf{SimGCL} \\ \footnotesize (SIGIR'22)} 
& \makecell[c]{\textbf{MBGCN} \\ \footnotesize (SIGIR'20)}   
& \makecell[c]{\textbf{MB-CGCN} \\ \footnotesize (TKDE'23)} 
& \makecell[c]{\textbf{PKEF} \\ \footnotesize (CIKM'23)}   
& \makecell[c]{\textbf{BCIPM} \\ \footnotesize (SIGIR'24)} 
& \makecell[c]{\textbf{S-MBRec} \\ \footnotesize (IJCAI'22)} 
& \makecell[c]{\textbf{MISSL} \\ \footnotesize (ICDE'24)} 
& \makecell[c]{\textbf{DeMBR} \\ \footnotesize (WSDM'25)} 
& \makecell[c]{\textbf{UIPL} \\ \footnotesize (TOIS'25)} 
& \makecell[c]{\textbf{MBID} \\ \footnotesize (CIKM'25)} 
& \textbf{\shortname} 
& \textbf{\textit{Imp.}} \\
\midrule

\multirow{4}{*}{\textbf{Taobao}}  
 & \textbf{HR@10}   
 & 0.0253 & 0.0400 & 0.0433 & 0.0989 & 0.1098 & \underline{0.1257} & 0.0568 & 0.0782 & 0.1083 & 0.1225 & 0.0784 
 & \textbf{0.1586*} & \textbf{26.17\%} \\
 & \textbf{NDCG@10} 
 & 0.0137 & 0.0236 & 0.0257 & 0.0471 & 0.0627 & 0.0708 & 0.0328 & 0.0629 & 0.0673 & \underline{0.0748} & 0.0412 
 & \textbf{0.0938*} & \textbf{25.40\%} \\
 & \textbf{HR@20}   
 & 0.0364 & 0.0522 & 0.0602 & 0.1368 & 0.1223 & \underline{0.1723} & 0.0891 & 0.1134 & 0.1427 & 0.1567 & 0.1172 
 & \textbf{0.1993*} & \textbf{15.67\%} \\
 & \textbf{NDCG@20} 
 & 0.0178 & 0.0265 & 0.0303 & 0.0595 & 0.0651 & 0.0833 & 0.0411 & 0.0759 & 0.0760 & \underline{0.0834} & 0.0497 
 & \textbf{0.1041*} & \textbf{24.82\%} \\
\midrule

\multirow{4}{*}{\textbf{TMall}}  
 & \textbf{HR@10}   
 & 0.0379 & 0.0709 & 0.0549 & 0.1083 & 0.1138 & 0.1401 & 0.0694 & 0.0691 & 0.1227 & 0.1422 & \textbf{0.1536} 
 & \underline{0.1468} & $-4.43\%$ \\
 & \textbf{NDCG@10} 
 & 0.0205 & 0.0399 & 0.0285 & 0.0426 & 0.0642 & 0.0738 & 0.0362 & 0.0552 & 0.0723 & \underline{0.0770} & 0.0655 
 & \textbf{0.0793*} & \textbf{2.99\%} \\
 & \textbf{HR@20}   
 & 0.0486 & 0.1018 & 0.0799 & 0.1370 & 0.1676 & 0.1903 & 0.1009 & 0.1012 & 0.1469 & 0.1932 & \underline{0.1965} 
 & \textbf{0.2023*} & \textbf{2.95\%} \\
 & \textbf{NDCG@20} 
 & 0.0231 & 0.0487 & 0.0345 & 0.0523 & 0.0562 & 0.0885 & 0.0438 & 0.0667 & 0.0782 & \underline{0.0900} & 0.0780 
 & \textbf{0.0929*} & \textbf{3.22\%} \\
\midrule

\multirow{4}{*}{\textbf{MovieLens-10M}}  
 & \textbf{HR@10}   
 & 0.0660 & 0.0695 & 0.0467 & 0.0619 & 0.0591 & 0.0782 & 0.0340 & 0.0415 & 0.0685 & \underline{0.0791} & 0.0327 
 & \textbf{0.0801*} & \textbf{1.26\%} \\
 & \textbf{NDCG@10} 
 & 0.0317 & 0.0338 & 0.0232 & 0.0292 & 0.0264 & 0.0362 & 0.0164 & 0.0198 & 0.0321 & \underline{0.0373} & 0.0218 
 & \textbf{0.0382*} & \textbf{2.41\%} \\
 & \textbf{HR@20}   
 & 0.1172 & 0.1228 & 0.0768 & 0.1062 & 0.0987 & 0.1345 & 0.0666 & 0.0768 & 0.1171 & \underline{0.1351} & 0.0550 
 & \textbf{0.1368*} & \textbf{1.26\%} \\
 & \textbf{NDCG@20} 
 & 0.0452 & 0.0478 & 0.0309 & 0.0402 & 0.0359 & 0.0501 & 0.0238 & 0.0282 & 0.0441 & \underline{0.0512} & 0.0356 
 & \textbf{0.0524*} & \textbf{2.34\%} \\
\bottomrule
\end{tabular}
}

\end{table*}

To validate the effectiveness of our model, we compare ~\shortname~ \\ with a range of state-of-the-art baselines across three real-world datasets. The results, summarized in Table~\ref{tab:main_result}, show that ~\shortname~ \\ achieves the best performance in most settings and consistently obtains the best NDCG results across all three datasets, demonstrating its effectiveness in modeling target-oriented preferences under heterogeneous behaviors. The key observations are as follows:

\begin{itemize}[leftmargin=0.3cm]

    \item   \textbf{Multi-behavior signals are essential for alleviating target sparsity.} Multi-behavior models significantly outperform single-behavior baselines relying solely on sparse purchases. 
    For example, on Taobao, the best multi-behavior baseline (BCIPM) exceeds the strongest single-behavior baseline (SimGCL) by \textbf{214\%} on HR@10, while on TMall, MBID improves over SimGCL by \textbf{116.6\%} on HR@10.
    These results highlight that auxiliary behaviors, such as \textit{view} and \textit{cart}, provide complementary supervision that cannot be replaced by augmenting sparse target behavior alone. 

    \item  \textbf{Denoising strategies are crucial for handling noisy and inconsistent auxiliary behaviors.} Standard multi-behavior models (e.g., MB-CGCN) lack effective noise filtering, leading to entangled perturbations in the embeddings. Denoising methods show clear advantages: on Taobao, DeMBR improves NDCG@10 by 42.9\% over MB-CGCN, while on TMall, UIPL achieves an 80.8\% improvement. This underscores the importance of noise suppression, as unfiltered spurious signals dominate naive fusion approaches and distort target preferences.

    \item \textbf{\shortname~ achieves strong overall performance through feature-frequency purification and reliability-aware fusion.} ~\shortname~ obtains the best results in most evaluation settings and consistently ranks first on all NDCG metrics. On Taobao, it improves over the strongest baseline by \textbf{26.17\%} on HR@10 and \textbf{25.40\%} on NDCG@10. On MovieLens-10M, it also achieves the best results across all four metrics. On TMall, although MBID obtains a higher HR@10, ~\shortname~ achieves the best NDCG@10, HR@20, and NDCG@20 results, indicating stronger ranking quality for top-ranked recommendations. These results support our design of purifying behavior-specific representations before global-context-guided fusion.
    
\end{itemize}

\subsection{Ablation Study}

%这里是一个图我们来看看
\begin{figure}

    \centering
    \includegraphics[width =\linewidth]{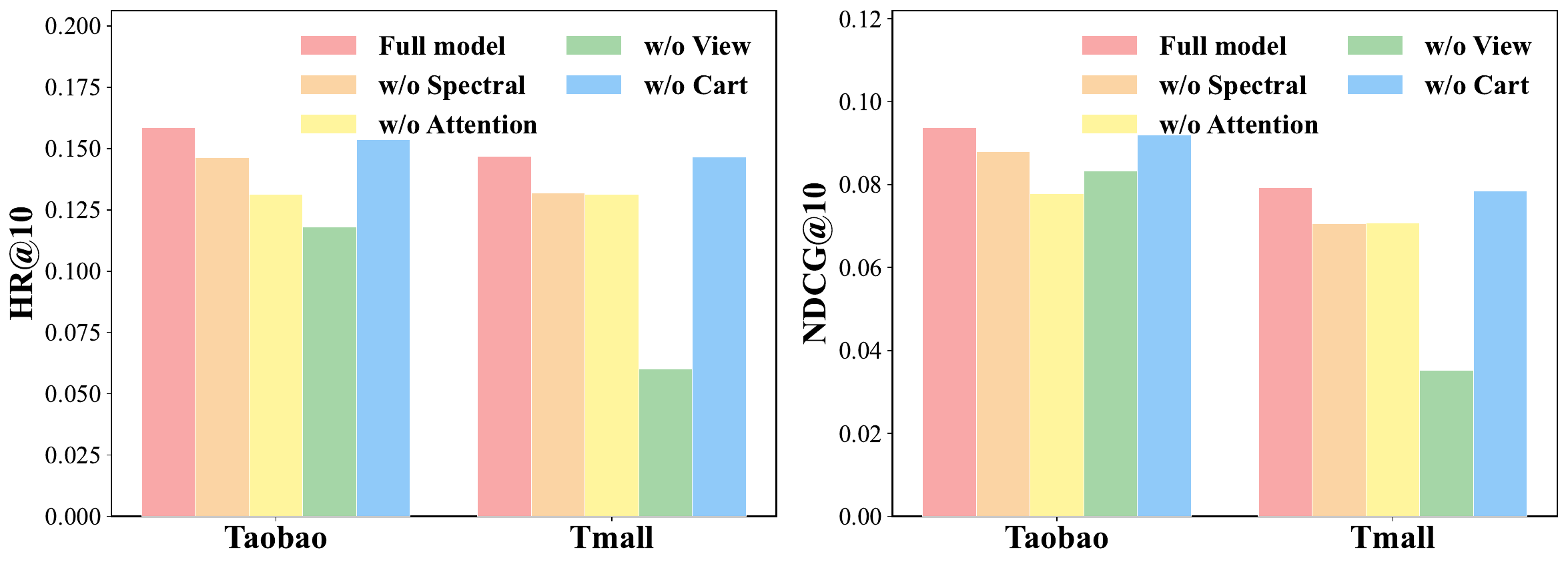}
    \caption{Ablation results of ~\shortname~on Taobao and TMall.}
    \label{fig:ablation}
    \vspace{-0.5cm}
\end{figure}

To evaluate the contribution of each component in ~\shortname, we construct the following variants: \textbf{w/o Spectral} removes \emph{Dynamic Feature-Level Spectral Filtering} and directly fuses the raw spatial embeddings; \textbf{w/o Attention} replaces \emph{Global-Context Attention Fusion} with mean pooling across behavior views; \textbf{w/o View} and \textbf{w/o Cart} remove the respective auxiliary interactions from training. The results on Taobao and TMall are presented in Figure~\ref{fig:ablation}. Due to the different behavior construction in MovieLens-10M (ratings are categorized into \emph{dislike/neutral/like}), we report its results in Appendix~\ref{app:ablation_ml}, which shows consistent trends.

\begin{itemize}[leftmargin=0.3cm]

    \item \textbf{Feature-frequency purification is essential for mitigating intra-behavior entanglement.} Removing spectral filtering consistently degrades performance. On \textbf{Taobao}, HR@10 and NDCG@10 decrease by \textbf{6.9\%} and \textbf{6.2\%}, respectively, while on \textbf{TMall}, the performance drop is even more significant, with HR@10 dropping by \textbf{10.0\%} and NDCG@10 by \textbf{10.9\%}. This confirms that after multi-hop propagation, perturbations from auxiliary behaviors become entangled with preference signals, highlighting the need for feature-frequency purification to suppress this entanglement without over-smoothing important signals. 

    \item \textbf{Global-context fusion is vital for calibrating inter-behavior reliability.} Replacing context-guided attention with simple mean pooling leads to a noticeable decline in performance. On \textbf{Taobao}, HR@10 and NDCG@10 drop by \textbf{16.8\%}, and on \textbf{TMall}, the performance decreases by \textbf{10.4\%} on HR@10 and \textbf{10.9\%} on NDCG@10. These results demonstrate that, when behavior reliability varies across users and contexts, using a unified fusion method makes the model more vulnerable to unreliable signals. In contrast, anchoring reliability estimation to a purified global context leads to more stable target prediction. 

    \item \textbf{Auxiliary behaviors are beneficial but require purification and calibration.} Removing the \textit{view} behavior causes the largest degradation, particularly on \textbf{TMall}, where HR@10 and NDCG@10 drop by \textbf{59.3\%} and \textbf{55.9\%}, respectively. This shows that high-coverage browsing signals play a crucial role in alleviating target sparsity. On the other hand, removing the \textit{cart} behavior has a minimal impact, suggesting that its contribution is either sparse or redundant given the other signals. 
    These findings highlight the importance of not just adding more behaviors but purifying each view to mitigate intra-view entanglement and calibrating cross-view reliability to avoid misaligned signals.

\end{itemize}

\subsection{Further Analysis}

\subsubsection{Impact of Dynamic Feature-Level Spectral Filtering}

To examine whether \emph{Dynamic Feature-Level Spectral Filtering} is necessary for mitigating \textit{intra-behavior representation entanglement}, we compare ~\shortname~ with three \emph{static} spectral baselines. \textbf{All-Pass} keeps the feature-frequency spectrum unchanged (i.e., no purification), while \textbf{Low-Pass} and \textbf{High-Pass} apply two complementary fixed masks that suppress different subsets of frequency bins. In contrast, ~\shortname~ learns \emph{view-adaptive} spectral modulation end-to-end under the target objective, enabling component-wise purification without imposing any fixed frequency prior.

Table~\ref{tab:pass} shows that ~\shortname~ consistently achieves the best results on both datasets. 
Notably, \textbf{All-Pass} causes a non-trivial drop (e.g., Taobao HR@10: $-7.82\%$), suggesting that merely transforming representations into a frequency parameterization is insufficient without active purification. 
Moreover, both \textbf{Low-Pass} and \textbf{High-Pass} lead to severe degradation (e.g., Taobao HR@10 decreases by over $37\%$), indicating that rigid band-selection is brittle and can easily suppress target-relevant cues or amplify perturbations. 

The superiority of ~\shortname~ over static filters confirms that learning view-adaptive patterns is essential. 
Furthermore, the significant performance gain achieved using a simple linear diagonal modulator validates our design choice: the Fast Fourier Transform (FFT) effectively disentangles complex spatial perturbations into separable frequency components. 
In this disentangled space, semantic signals and noise are sufficiently distinct that a parameter-efficient linear scaling suffices to suppress noise. 
This suggests that the core challenge in multi-behavior denoising is the \textit{representation basis} (solved by FFT) rather than the complexity of the filtering operation itself.
To further quantify this purification effect, we conduct frequency-domain visualization and information-theoretic analysis. Due to space limitations, detailed results are deferred to Appendix~\ref{sec:appendix_spectral}, where we visualize behavior-dependent spectral modulation patterns and examine the target alignment of the learned representations using mutual information.

\begin{table}[t]
\centering
\caption{Comparison of spectral filtering strategies.}
\label{tab:pass}
\resizebox{\linewidth}{!}{
\begin{tabular}{l|cc|cc|cc}
\toprule
\multirow{2}{*}{\textbf{Strategy}} 
& \multicolumn{2}{c|}{\textbf{Taobao}} 
& \multicolumn{2}{c|}{\textbf{TMall}} 
& \multicolumn{2}{c}{\textbf{MovieLens-10M}} \\
\cmidrule(lr){2-3}\cmidrule(lr){4-5}\cmidrule(lr){6-7}
& \textbf{HR@10} & \textbf{NDCG@10}
& \textbf{HR@10} & \textbf{NDCG@10}
& \textbf{HR@10} & \textbf{NDCG@10} \\
\midrule
\textbf{\shortname} 
& \textbf{0.1586*} & \textbf{0.0938*}
& \textbf{0.1468*} & \textbf{0.0793*}
& \textbf{0.0801*} & \textbf{0.0382*} \\
\midrule
All-Pass 
& 0.1462 \small{(-7.8\%)} & 0.0879 \small{(-6.3\%)}
& 0.1318 \small{(-10.2\%)} & 0.0706 \small{(-11.0\%)}
& 0.0727 \small{(-9.2\%)} & 0.0349 \small{(-8.5\%)} \\
Low-Pass 
& 0.0930 \small{(-41.4\%)} & 0.0568 \small{(-39.5\%)}
& 0.1026 \small{(-30.1\%)} & 0.0555 \small{(-30.0\%)}
& 0.0517 \small{(-35.4\%)} & 0.0248 \small{(-35.2\%)} \\
High-Pass 
& 0.0985 \small{(-37.9\%)} & 0.0595 \small{(-36.6\%)}
& 0.1031 \small{(-29.8\%)} & 0.0564 \small{(-28.9\%)}
& 0.0535 \small{(-33.3\%)} & 0.0256 \small{(-33.1\%)} \\
\bottomrule
\end{tabular}
}

\end{table}

\subsubsection{Robustness against Interaction Noise.}

\begin{figure}
    \centering
    \includegraphics[width =\linewidth]{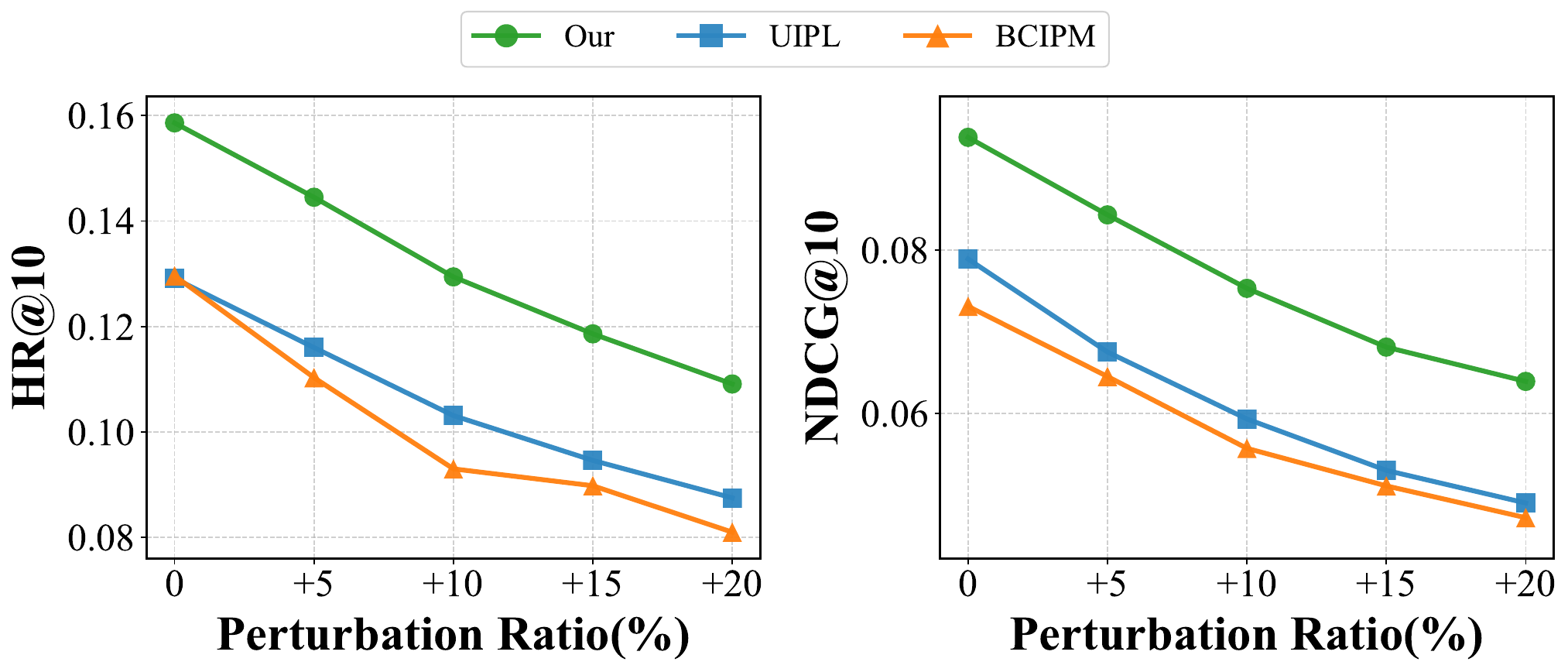}
    \caption{Performance comparison under varying levels of random interaction noise on the Taobao dataset.}
    \label{fig:robustness}
    \vspace{-0.5cm}
\end{figure}

To evaluate the robustness of ~\shortname~ in noisy environments, we injected \textbf{random edge addition noise} into the interaction graph. Specifically, we randomly sampled non-interacted user-item pairs and added them as spurious edges to the training graph, varying the noise ratio from $0\%$ to $20\%$ of the original interactions. This setup simulates real-world noisy feedback and directly targets two key challenges addressed by our model: (1) As noise propagates through multi-hop aggregation, it exacerbates intra-behavior entanglement by mixing spurious signals with genuine preferences; (2) The distorted auxiliary interactions amplify inter-behavior inconsistency, thereby testing the model's ability to prevent misleading signals from dominating the fusion process. 

We compare ~\shortname~ with denoising baselines \textbf{UIPL} and \textbf{BCIPM} on Taobao. As shown in Figure~\ref{fig:robustness}, ~\shortname~ consistently outperforms baselines, exhibiting superior robustness under increasing noise. While baselines degrade rapidly (up to $20\%$ corruption), ~\shortname~ maintains stable performance, preserving ranking quality. This advantage stems from suppressing perturbations via dynamic spectral filtering, rather than relying on spatial propagation which struggles with noisy edges.

Additionally, global-context fusion and the residual backbone provide a stable structural anchor, preventing noisy auxiliary behaviors from dominating the representation. Consequently, ~\shortname~ maintains robust target ranking even under significant interaction noise.

\subsubsection{Case Study: What Noise Is Suppressed?}

\begin{table}[t]
\centering
\caption{Effect of spectral purification on recommendation popularity on TMall.}
\label{tab:popularity_case}
\small
\setlength{\tabcolsep}{5pt}
\begin{tabular}{lccc}
\toprule
\textbf{View} & \textbf{Raw} & \textbf{Clean} & \textbf{Drop} \\
\midrule
View   & 103.68 & 41.69 & 59.8\% \\
Cart   & 72.26  & 51.87 & 28.2\% \\
Purchase    & 74.38  & 39.03 & 47.5\% \\
Global & 104.87 & 42.65 & 59.3\% \\
\bottomrule
\end{tabular}
\vspace{-0.5cm}
\end{table}

To further interpret the effect of spectral purification, we examine whether it reduces excessive popularity concentration in the resulting recommendation lists. In graph-based recommendation, noisy propagation may amplify signals associated with frequently interacted items, causing the learned representations to overly favor popularity-dominated recommendations. We therefore use the change in recommended-item popularity as an observable indicator of the denoising effect, rather than assuming that popular items are inherently noisy.

Specifically, on the TMall dataset, we use the raw and purified representations of each view to independently generate Top-20 recommendation lists through inner-product scoring. We then compute the average popularity of recommended items, measured by their historical interaction frequency in the training set. As shown in Table~\ref{tab:popularity_case}, spectral purification consistently reduces the average popularity across all examined views, with decreases ranging from 28.2\% to 59.8\%. In particular, the substantial reductions on the View and Global representations suggest that ~\shortname~ effectively alleviates popularity-dominated signals amplified during propagation, allowing more specific target-relevant preference evidence to contribute to the recommendation.

\subsubsection{Impact of Residual Hyperparameter $\alpha$.}

\begin{figure}[t]
    \centering
    \includegraphics[width =0.9\linewidth]{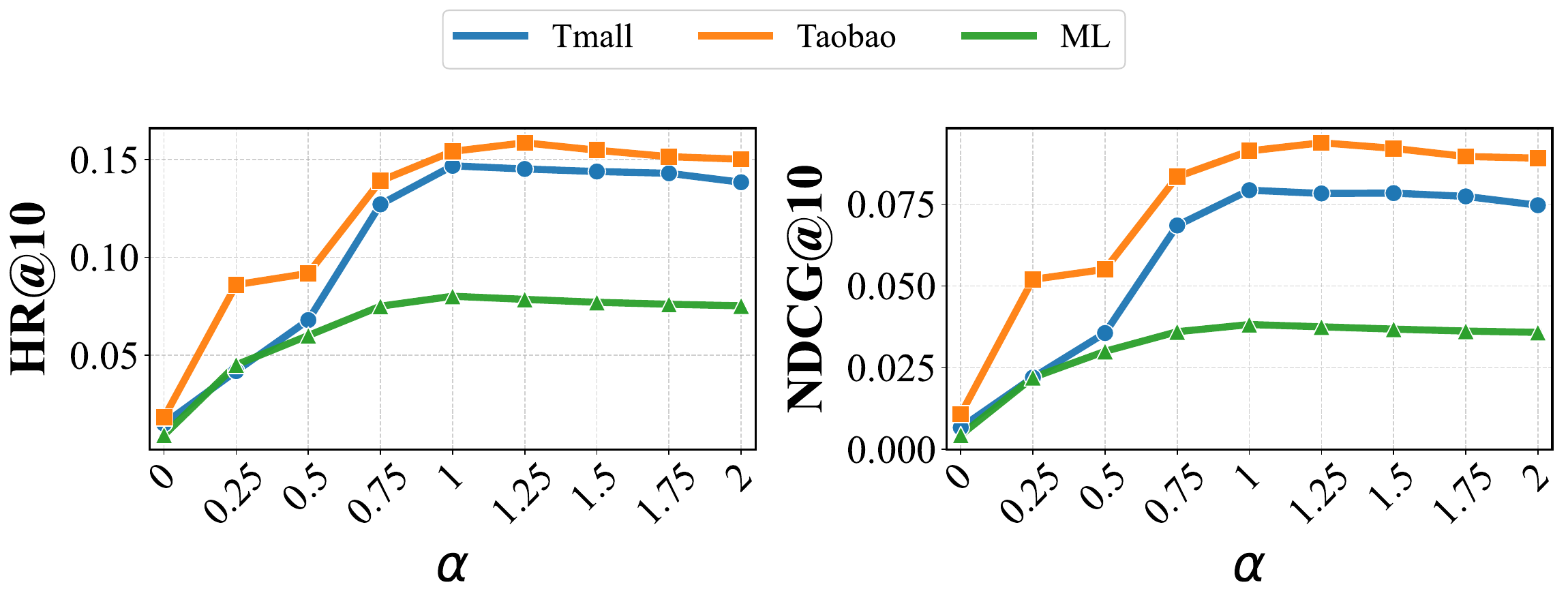}
    \caption{Performance impact of the residual hyperparameter $\alpha$ on different datasets.}
    \label{fig:alpha_sensitivity}
    \vspace{-0.5cm}
\end{figure}

The residual coefficient $\alpha$ is a crucial hyperparameter in ~\shortname, determining the contribution of the global-graph residual branch to the final representation. This design reflects our core principle that robust multi-behavior recommendation requires a balance between effective noise suppression and the preservation of reliable collaborative structure. While dynamic spectral filtering addresses behavior-dependent perturbations, the residual branch retains a stable global context essential for preference modeling. We vary $\alpha$ from $0$ to $2$ in increments of $0.25$, and present the results in Figure~\ref{fig:alpha_sensitivity}.

As shown in Figure~\ref{fig:alpha_sensitivity}, performance improves with increasing $\alpha$, peaking at moderate values. When $\alpha$ is near zero, the model relies predominantly on purified spectral signals, but the absence of structural context degrades ranking quality. As $\alpha$ increases, the introduction of global structural information enhances prediction accuracy, emphasizing the residual branch's role in complementing spectral denoising. However, when $\alpha$ becomes too large, performance declines, suggesting that excessive reliance on raw spatial information reintroduces noisy correlations, undermining the benefits of dynamic spectral filtering. 
These findings confirm that the residual connection acts as a structural stabilizer. The sensitivity of $\alpha$ indicates that the appropriate contribution of the residual global branch is dataset-dependent, requiring a balance between purified behavior-specific signals and global collaborative structure. 

%这里是一个图我们来看看

\section{Conclusion}

In this paper, we shifted the perspective from the spatial domain to the \textit{feature-frequency domain}, introducing ~\shortname~ to address the limitations of traditional spatial-only denoising methods in multi-behavior recommendation. The core challenges we tackled were the complex entanglement of behavior-specific signals and the difficulty in distinguishing between stochastic noise and informative niche signals in the spatial view. To mitigate intra-behavior entanglement, we introduced the \textbf{Dynamic Feature-Level Spectral Filtering} module, which adaptively modulated frequency components along the embedding dimension, effectively suppressing random fluctuations while preserving essential preference patterns. To address inter-behavior reliability heterogeneity, we proposed Global-Context Attention Fusion, which uses the purified global representation as a context anchor to perform reliability-aware aggregation over behavior-specific representations. 
 Our extensive experiments on three real-world datasets demonstrate that ~\shortname~ achieves the best results in most evaluation settings, including consistently superior NDCG performance across all datasets, while also exhibiting strong robustness against interaction noise. These findings validate the effectiveness of feature-frequency purification and global-context-guided fusion for handling heterogeneous auxiliary behaviors in multi-behavior recommendation.

\begin{acks}
This work was supported in part by grants from the National Natural Science Foundation of China (Grant No. U25A20445, 62272254).
\end{acks}

\newpage

\bibliographystyle{ACM-Reference-Format}

\bibliography{sample-base}

\appendix

\vspace{-0.2cm}
\section{Detailed Description of ~\shortname~ Algorithm}
\label{appendix:algorithm}

Algorithm~\ref{alg:shortname} outlines the end-to-end training procedure of ~\shortname. The framework is optimized via mini-batch stochastic gradient descent. 
In each iteration, the model sequentially performs multi-view graph propagation, feature-level spectral purification, and global-context attention fusion to distill robust representations. 
Finally, the parameters are updated via backpropagation derived strictly from the target behavior loss.

\section{Ablation Study on MovieLens-10M}
\label{app:ablation_ml}

\begin{figure}
    \centering
    
    \includegraphics[width =\linewidth]{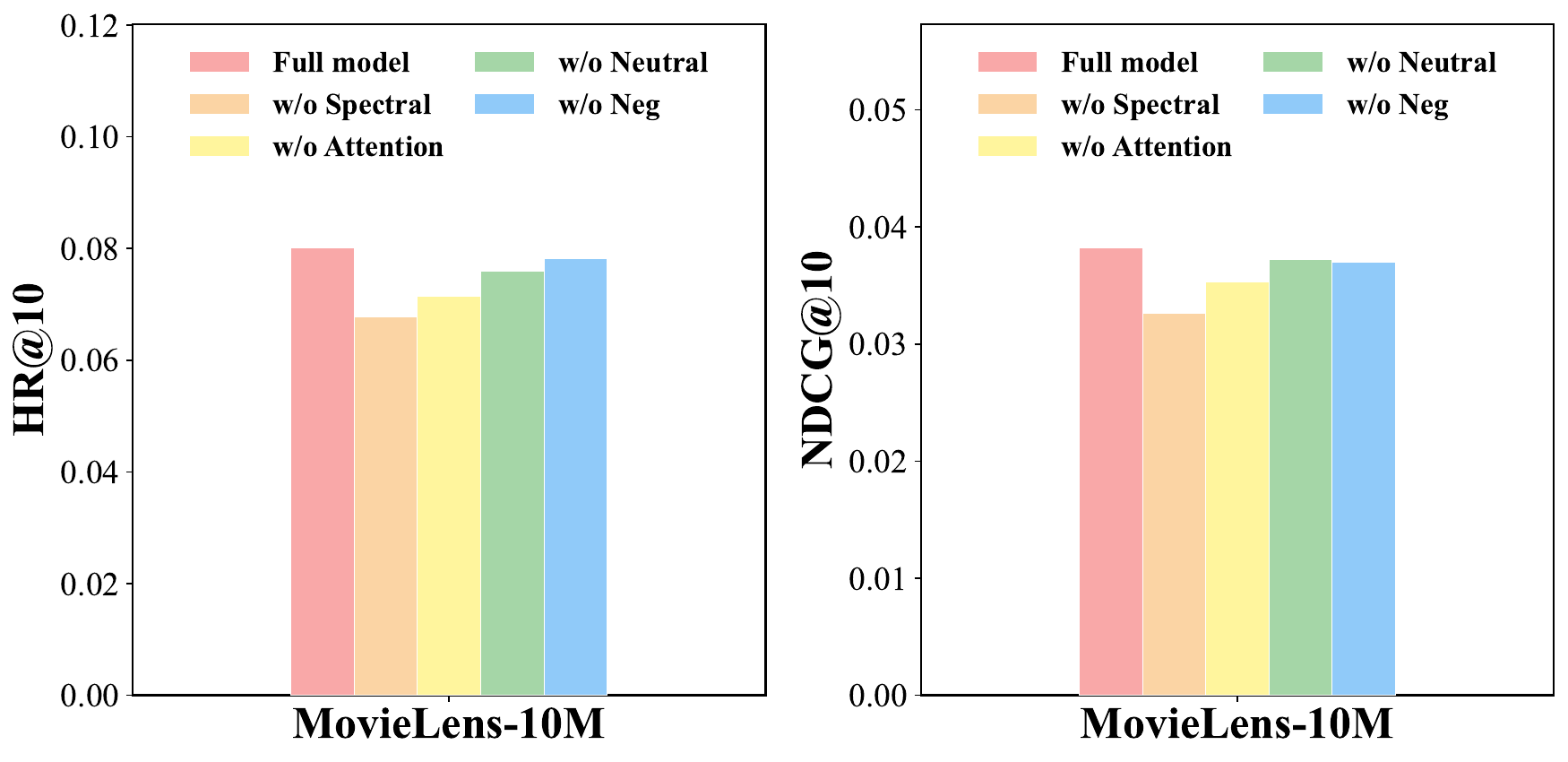}
    \caption{Ablation study on the MovieLens-10M dataset, evaluating the impact of different components of ~\shortname.}
    \label{fig:ablation_ml}

\end{figure}

Unlike the e-commerce datasets (Taobao/TMall), MovieLens-10M
constructs implicit behaviors by converting explicit ratings into
\textit{Like} (target), \textit{Neutral}, and \textit{Dislike}.
Figure~\ref{fig:ablation_ml} reports the performance of ~\shortname~
and its variants in this rating-based multi-behavior setting.

Consistent with the findings in the main text, removing the core
components causes the most significant performance degradation.
Specifically, removing \textbf{Dynamic Spectral Filtering}
(\textbf{w/o Spectral}) results in the lowest HR@10 and NDCG@10,
demonstrating the effectiveness of feature-frequency purification
for handling entangled preference and perturbation signals in
rating-derived behaviors. Replacing \textbf{Global-Context Attention
Fusion} with mean pooling (\textbf{w/o Attention}) also causes a
marked decline, confirming the importance of global-context-guided
reliability calibration for heterogeneous behaviors.

We further examine the contribution of different auxiliary behaviors.
Removing \textbf{Neutral} interactions (\textbf{w/o Neutral}) leads
to a noticeable performance drop, suggesting that neutral feedback
provides useful contextual information for predicting the target
\textit{Like} behavior. In comparison, removing \textbf{Dislike}
interactions (\textbf{w/o Neg}) has a smaller impact, indicating that
\textit{Neutral} feedback serves as a more informative auxiliary signal
in this dataset. Overall, the results on MovieLens-10M are consistent
with those on Taobao and TMall, demonstrating the generalizability of
~\shortname~ across different multi-behavior construction settings.

\begin{algorithm}[t]
\renewcommand{\algorithmicrequire}{\textbf{Input:}}
\renewcommand{\algorithmicensure}{\textbf{Output:}}
\caption{\small{The Algorithm of ~\shortname.}}
\label{alg:shortname}
\begin{algorithmic}[1]
\REQUIRE Training interactions $\{\mathbf{Y}_b\}_{b\in\mathcal{B}}$, target behavior $target$, epochs $T$, learning rate $\eta$, regularization $\lambda$, residual weight $\alpha$;
\ENSURE Trained parameters $\Theta$;

\STATE Construct $\{\mathcal{G}_b\}_{b\in\mathcal{B}}$ and $\mathcal{G}_{global}$ (training only);
\STATE Initialize node embeddings $\mathbf{E}^{(0)}$ and model parameters $\Theta$;

\FOR{$epoch=1,2,\ldots,T$}
    \STATE Sample a mini-batch of target triplets $(u,i,j)$;

    \STATE \textbf{Step 1: Multi-View Graph Propagation}
    \STATE Obtain $\mathbf{E}^{raw}_{global}$ and $\{\mathbf{E}^{raw}_b\}_{b\in\mathcal{B}}$ by Eqs.~(\ref{eq:gcn_prop})--(\ref{eq:gcn_readout});

    \STATE \textbf{Step 2: Dynamic Feature-Level Spectral Filtering}
    \STATE Transform to feature-frequency domain by Eq.~(\ref{eq:rfft});
    \STATE Apply behavior-adaptive spectral modulation by Eq.~(\ref{eq:spec_mod});
    \STATE Reconstruct purified embeddings by Eq.~(\ref{eq:irfft}), yielding $\mathbf{E}^{clean}_{global}$ and $\{\mathbf{E}^{clean}_b\}$;

    \STATE \textbf{Step 3: Global-Context Attention Fusion}
    \STATE Compute compatibility $\{\mathbf{S}_b\}$ by Eq.~(\ref{eq:compat});
    \STATE Compute attention weights $\{\mathbf{A}_b\}$ by Eq.~(\ref{eq:attn});
    \STATE Fuse multi-behavior representations by Eq.~(\ref{eq:fusion}) to obtain $\mathbf{E}^{final}$;

    \STATE \textbf{Step 4: Target Behavior Optimization}
    \STATE Compute scores $\hat{y}_{ui},\hat{y}_{uj}$ by Eq.~(\ref{eq:pred});
    \STATE Compute loss $\mathcal{L}$ by Eq.~(\ref{eq:bpr}) and update $\Theta \leftarrow \Theta - \eta \nabla_{\Theta}\mathcal{L}$;

    \IF{early stopping is met} \STATE \textbf{break}; \ENDIF
\ENDFOR

\STATE \textbf{return} trained parameters $\Theta$;
\end{algorithmic}
\end{algorithm}

\section{Empirical Analysis of Spectral Purification}
\label{sec:appendix_spectral}

\subsection{Visualization of Spectral Modulation}

\begin{figure}
    \centering
    \includegraphics[width =\linewidth]{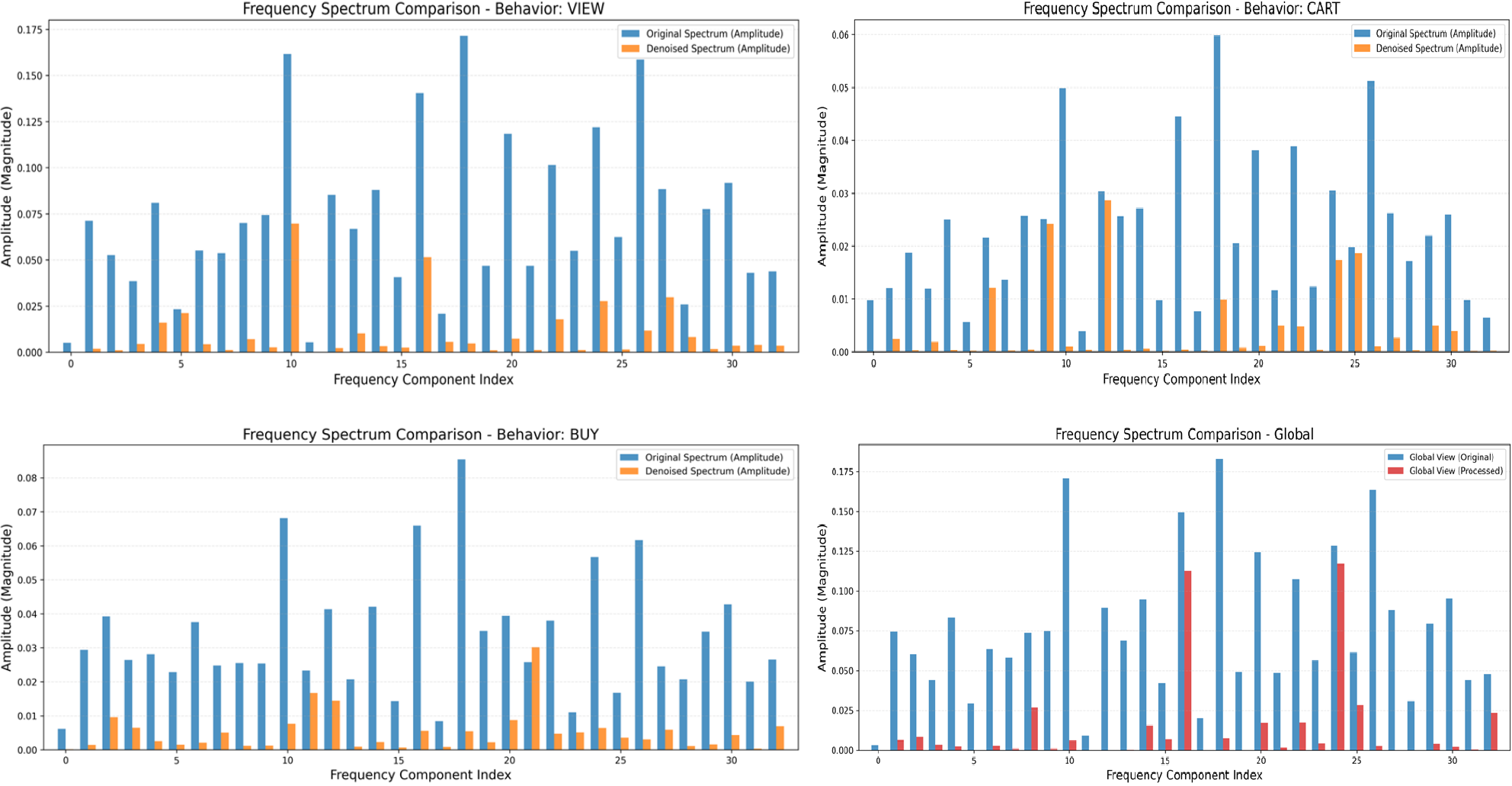}
    \caption{Visualization of Spectral Purification across different behaviors.}
    \label{fig:spectrum_vis}

\end{figure}

To gain intuitive insight into how~\shortname~ mitigates intra-behavior entanglement, we conduct a frequency-domain visualization of the user embeddings in Figure~\ref{fig:spectrum_vis}. 
The figure presents a comparative spectral analysis between the raw embeddings derived from spatial aggregation ($E^{raw}$) and the purified embeddings after spectral filtering ($E^{clean}$) across different behaviors.

A key observation is that the raw embeddings consistently exhibit high energy uniformly distributed across the entire spectrum. This implies severe representation entanglement, where noise is indistinguishable from true preference signals. 
After modulation, ~\shortname~ displays distinct, \textit{behavior-adaptive} filtering patterns that align with the semantic nature of each interaction type.
For exploratory interactions like \textit{'View'}, which are often fleeting and contain frequent stochastic state changes, the spectrum manifests as dominant \textbf{high-frequency noise}. 
Consequently, ~\shortname~ aggressively suppresses the high-frequency bands in the 'View' subplot (where Orange bars are significantly lower than Blue ones), effectively functioning as a \textbf{``semantic low-pass filter''} to strip away incidental interaction noise.
\textit{In stark contrast}, intent-driven behaviors such as \textit{'Cart'} and \textit{'Purchase'} reflect stable user interests corresponding to low-frequency components, yet they also contain specific high-frequency details representing niche preferences. 
For these behaviors, the modulation exhibits a \textbf{selective preservation strategy}: it retains not only the dominant low-frequency core but also specific high-frequency peaks necessary for distinguishing fine-grained interests.

\subsection{Information-Theoretic Validation}

\begin{figure}
    \centering
    \includegraphics[width =\linewidth]{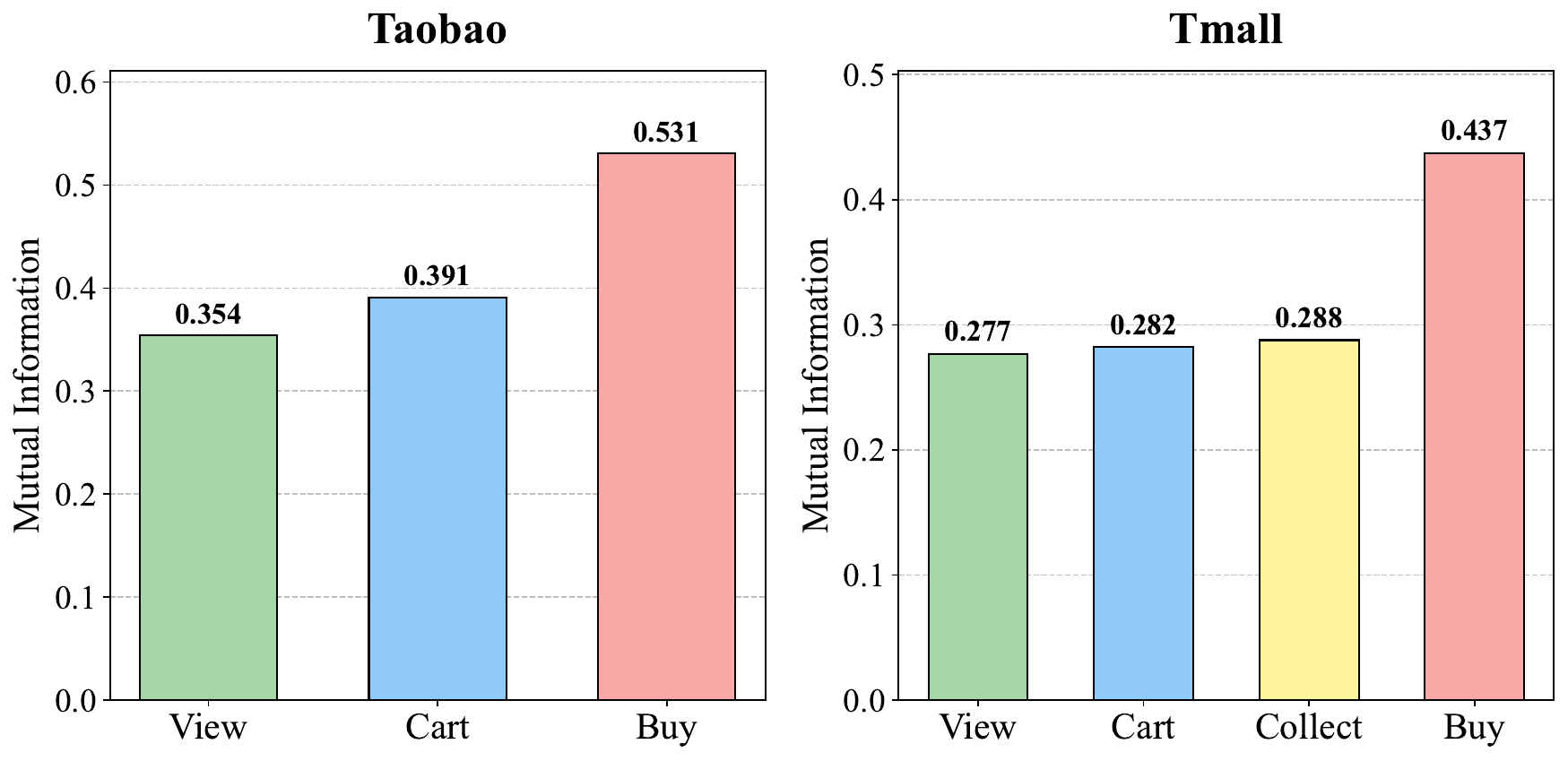}
    \caption{Information-Theoretic Validation of Target Alignment.}
    \label{fig:mutual_info}

\end{figure}

To quantitatively verify that our spectral filtering mechanism preserves ``informative preferences'' rather than retaining ``stochastic noise,'' we employ Mutual Information (MI) to measure the dependency between the learned user representations and the ground-truth labels.
As illustrated in Figure~\ref{fig:mutual_info}, we observe a distinct \textit{information hierarchy} across both datasets. Specifically, the learned representations exhibit the strongest dependency with the target behavior (\textbf{Buy}), which consistently achieves the highest MI scores on both Taobao and TMall. This stands in sharp contrast to the auxiliary behaviors (e.g., View and Cart), which show significantly lower mutual information despite their significantly higher interaction frequency.

Crucially, the lower MI for \textit{View} does not imply that browsing behaviors are ignored. 
Instead, it indicates that ~\shortname~ effectively resolves the \textit{Volume-Value Paradox}. 
While \textit{View} data provides the most abundant source of context (essential for mitigating sparsity), it is inherently noisy and high-entropy. 
Our spectral filtering acts as a \textit{semantic sieve}: it filters out the stochastic, incidental views that do not translate to purchases, while selectively retaining the latent interest patterns embedded within browsing histories. 
Therefore, the moderate MI score for \textit{View} reflects a healthy state of \textit{selective utilization}—the model leverages auxiliary data for necessary context without overfitting to its noisy distribution.

This finding confirms the robustness of our approach against data skew. 
A naive model, overwhelmed by the volume of auxiliary interactions, would likely be dominated by high-frequency signals, resulting in a representation that mirrors the noisy input (high MI with View). 
The fact that ~\shortname~ maintains maximal alignment with the sparse target behavior provides compelling empirical evidence that our method successfully suppresses incidental noise. 
Consequently, the distilled representations remain robust and specifically oriented towards the core target objective.

\subsection{Spectral versus Spatial Modulation}
\label{sec:spatial_alternatives}

To further examine whether the effectiveness of Dynamic Feature-Level Spectral Filtering mainly comes from introducing additional learnable parameters, we replace the spectral filtering module with two spatial-domain alternatives while keeping the remaining model structure unchanged. Specifically, \emph{w/ Spatial Gate} applies a learnable per-dimension gate directly to the raw spatial embeddings, while \emph{w/ Spatial MLP} employs a lightweight nonlinear transformation in the original embedding space. These variants introduce adaptive feature transformations without performing the RFFT-based feature-frequency modulation used in ~\shortname. 

\begin{table}[t]
\centering
\caption{Comparison with spatial-domain alternatives.}
\label{tab:spatial_alternatives}
\small
\setlength{\tabcolsep}{3.5pt}
\resizebox{\columnwidth}{!}{
\begin{tabular}{llcccc}
\toprule
\textbf{Dataset} & \textbf{Variant} &
\textbf{HR@10} & \textbf{NDCG@10} &
\textbf{HR@20} & \textbf{NDCG@20} \\
\midrule
\multirow{3}{*}{Taobao}
& \textbf{\shortname}    & \textbf{0.1586} & \textbf{0.0938} & \textbf{0.1993} & \textbf{0.1041} \\
& w/ Spatial Gate        & 0.1371 & 0.0819 & 0.1764 & 0.0918 \\
& w/ Spatial MLP         & 0.1103 & 0.0655 & 0.1434 & 0.0739 \\
\midrule
\multirow{3}{*}{TMall}
& \textbf{\shortname}    & \textbf{0.1468} & \textbf{0.0793} & \textbf{0.2023} & \textbf{0.0929} \\
& w/ Spatial Gate        & 0.1373 & 0.0725 & 0.1908 & 0.0855 \\
& w/ Spatial MLP         & 0.0957 & 0.0503 & 0.1378 & 0.0606 \\
\bottomrule
\end{tabular}
}

\end{table}

As shown in Table~\ref{tab:spatial_alternatives}, replacing the spectral filtering module with either spatial-domain alternative consistently degrades performance on both datasets. Compared with \emph{w/ Spatial Gate}, ~\shortname~ improves HR@10 by 15.7\% on Taobao and 6.9\% on TMall, while the improvements over \emph{w/ Spatial MLP} reach 43.8\% and 53.4\%, respectively. Similar gains are observed on NDCG@10, HR@20, and NDCG@20. These results indicate that the effectiveness of ~\shortname~ does not merely arise from introducing additional adaptive transformations in the original embedding space. Instead, performing target-supervised modulation in the feature-frequency space provides a more effective representation basis for suppressing perturbation-dominated components and preserving target-relevant signals.

\end{document}